\input harvmac
\input epsf

\newcount\figno
\figno=0
\def\fig#1#2{
\par\begingroup\parindent=0pt\leftskip=1cm\rightskip=1cm\parindent=0pt
\baselineskip=11pt
\global\advance\figno by 1
\midinsert
\centerline{\epsfbox{#2}}
\vskip 12pt
{\bf Fig.\ \the\figno: } #1\par
\endinsert\endgroup\par
}
\def\figlabel#1{\xdef#1{\the\figno}}

\def\Li{\mathop{\rm Li}\nolimits}

\nref\GreenSW{
  M.~B.~Green, J.~H.~Schwarz and L.~Brink,
  ``${\cal N}=4$ Yang-Mills And ${\cal N}=8$
  Supergravity As Limits Of String Theories,''
  Nucl.\ Phys.\  B {\bf 198}, 474 (1982).
}

\nref\qgraf{
  P.~Nogueira,
  ``Automatic Feynman graph generation,''
  J.\ Comput.\ Phys.\  {\bf 105}, 279 (1993).
}

\nref\UsyukinaJD{
  N.~I.~Usyukina and A.~I.~Davydychev,
  ``An Approach to the evaluation of three and four point ladder diagrams,''
  Phys.\ Lett.\  B {\bf 298}, 363 (1993).
}

\nref\UsyukinaCH{
  N.~I.~Usyukina and A.~I.~Davydychev,
  ``Exact results for three and four point ladder diagrams with an arbitrary
  number of rungs,''
  Phys.\ Lett.\  B {\bf 305}, 136 (1993).
}

\nref\BroadhurstIB{
  D.~J.~Broadhurst,
  ``Summation of an infinite series of ladder diagrams,''
  Phys.\ Lett.\  B {\bf 307}, 132 (1993).
}

\nref\BernZX{
  Z.~Bern, L.~J.~Dixon, D.~C.~Dunbar and D.~A.~Kosower,
  ``One loop $n$
  point gauge theory amplitudes, unitarity and collinear limits,''
  Nucl.\ Phys.\  B {\bf 425}, 217 (1994)
  [arXiv:hep-ph/9403226].
}

\nref\BernCG{
  Z.~Bern, L.~J.~Dixon, D.~C.~Dunbar and D.~A.~Kosower,
  ``Fusing gauge theory tree amplitudes into loop amplitudes,''
  Nucl.\ Phys.\  B {\bf 435}, 59 (1995)
  [arXiv:hep-ph/9409265].
}

\nref\BernNH{
  Z.~Bern, J.~S.~Rozowsky and B.~Yan,
  ``Two-loop four-gluon amplitudes in ${\cal N} = 4$ super-Yang-Mills,''
  Phys.\ Lett.\  B {\bf 401}, 273 (1997)
  [arXiv:hep-ph/9702424].
}

\nref\BernSC{
  Z.~Bern, L.~J.~Dixon and D.~A.~Kosower,
  ``One-loop amplitudes for $e^+ e^-$ to four partons,''
  Nucl.\ Phys.\  B {\bf 513}, 3 (1998)
  [arXiv:hep-ph/9708239].
}

\nref\SmirnovGC{
  V.~A.~Smirnov,
  ``Analytical result for dimensionally regularized massless on-shell double
  box,''
  Phys.\ Lett.\  B {\bf 460}, 397 (1999)
  [arXiv:hep-ph/9905323].
}

\nref\TauskVH{
  J.~B.~Tausk,
  ``Non-planar massless two-loop Feynman diagrams with four on-shell legs,''
  Phys.\ Lett.\  B {\bf 469}, 225 (1999)
  [arXiv:hep-ph/9909506].
}

\nref\KajantieHV{
  K.~Kajantie, M.~Laine and Y.~Schroder,
 ``A simple way to generate high order vacuum graphs,''
  Phys.\ Rev.\  D {\bf 65}, 045008 (2002)
  [arXiv:hep-ph/0109100].
}

\nref\IsaevTK{
  A.~P.~Isaev,
  ``Multi-loop Feynman integrals and conformal quantum mechanics,''
  Nucl.\ Phys.\  B {\bf 662}, 461 (2003)
  [arXiv:hep-th/0303056].
}

\nref\AnastasiouKJ{
  C.~Anastasiou, Z.~Bern, L.~J.~Dixon and D.~A.~Kosower,
  ``Planar amplitudes in maximally supersymmetric Yang-Mills theory,''
  Phys.\ Rev.\ Lett.\  {\bf 91}, 251602 (2003)
  [arXiv:hep-th/0309040].
}

\nref\HahnFE{
  T.~Hahn,
  ``CUBA: A library for multidimensional numerical integration,''
  Comput.\ Phys.\ Commun.\  {\bf 168}, 78 (2005)
  [arXiv:hep-ph/0404043].
}

\nref\BrittoNC{
  R.~Britto, F.~Cachazo and B.~Feng,
  ``Generalized unitarity and one-loop amplitudes in
  ${\cal N} = 4$ super-Yang-Mills,''
  Nucl.\ Phys.\  B {\bf 725}, 275 (2005)
  [arXiv:hep-th/0412103].
}

\nref\BernIZ{
  Z.~Bern, L.~J.~Dixon and V.~A.~Smirnov,
  ``Iteration of planar amplitudes in maximally supersymmetric Yang-Mills
  theory at three loops and beyond,''
  Phys.\ Rev.\  D {\bf 72}, 085001 (2005)
  [arXiv:hep-th/0505205].
} 

\nref\BuchbinderWP{
  E.~I.~Buchbinder and F.~Cachazo,
  ``Two-loop amplitudes of gluons and octa-cuts in
  ${\cal N} = 4$ super Yang-Mills,''
  JHEP {\bf 0511}, 036 (2005)
  [arXiv:hep-th/0506126].
}

\nref\CzakonRK{
  M.~Czakon,
  ``Automatized analytic continuation of Mellin-Barnes integrals,''
  Comput.\ Phys.\ Commun.\  {\bf 175}, 559 (2006)
  [arXiv:hep-ph/0511200].
}

\nref\CachazoMQ{
  F.~Cachazo, M.~Spradlin and A.~Volovich,
  ``Hidden beauty in multiloop amplitudes,''
  JHEP {\bf 0607}, 007 (2006)
  [arXiv:hep-th/0601031].
}

\nref\CachazoTJ{
  F.~Cachazo, M.~Spradlin and A.~Volovich,
  ``Iterative structure within the five-particle two-loop amplitude,''
  Phys.\ Rev.\  D {\bf 74}, 045020 (2006)
  [arXiv:hep-th/0602228].
}

\nref\BernVW{
  Z.~Bern, M.~Czakon, D.~A.~Kosower, R.~Roiban and V.~A.~Smirnov,
  ``Two-loop iteration of five-point ${\cal N} = 4$
  super-Yang-Mills amplitudes,''
  Phys.\ Rev.\ Lett.\  {\bf 97}, 181601 (2006)
  [arXiv:hep-th/0604074].
}

\nref\DrummondRZ{
  J.~M.~Drummond, J.~Henn, V.~A.~Smirnov and E.~Sokatchev,
  ``Magic identities for conformal four-point integrals,''
  JHEP {\bf 0701}, 064 (2007)
  [arXiv:hep-th/0607160].
}

\nref\BernEW{
  Z.~Bern, M.~Czakon, L.~J.~Dixon, D.~A.~Kosower and V.~A.~Smirnov,
  ``The Four-Loop Planar Amplitude and Cusp Anomalous Dimension in Maximally
  Supersymmetric Yang-Mills Theory,''
  Phys.\ Rev.\  D {\bf 75}, 085010 (2007)
  [arXiv:hep-th/0610248].
}

\nref\CachazoAZ{
  F.~Cachazo, M.~Spradlin and A.~Volovich,
  ``Four-Loop Cusp Anomalous Dimension From Obstructions,''
  Phys.\ Rev.\  D {\bf 75}, 105011 (2007)
  [arXiv:hep-th/0612309].
}

\nref\BernCT{
  Z.~Bern, J.~J.~M.~Carrasco, H.~Johansson and D.~A.~Kosower,
  ``Maximally supersymmetric planar Yang-Mills amplitudes at five loops,''
  arXiv:0705.1864 [hep-th].
}

\nref\AldayHR{
  L.~F.~Alday and J.~Maldacena,
  ``Gluon scattering amplitudes at strong coupling,''
  JHEP {\bf 0706}, 064 (2007)
  [arXiv:0705.0303 [hep-th]].
}

\nref\BuchbinderHM{
  E.~I.~Buchbinder,
  ``Infrared Limit of Gluon Amplitudes at Strong Coupling,''
  Phys.\ Lett.\  B {\bf 654}, 46 (2007)
  [arXiv:0706.2015 [hep-th]].
}

\nref\DrummondAU{
  J.~M.~Drummond, G.~P.~Korchemsky and E.~Sokatchev,
  ``Conformal properties of four-gluon planar amplitudes and Wilson loops,''
  arXiv:0707.0243 [hep-th].
}

\nref\BrandhuberYX{
  A.~Brandhuber, P.~Heslop and G.~Travaglini,
  ``MHV Amplitudes in ${\cal N}=4$ Super Yang-Mills and Wilson Loops,''
  arXiv:0707.1153 [hep-th].
}

\nref\CachazoAD{
  F.~Cachazo, M.~Spradlin and A.~Volovich,
  ``Four-Loop Collinear Anomalous Dimension in ${\cal N} = 4$
  Yang-Mills Theory,''
  arXiv:0707.1903 [hep-th].
}

\nref\KruczenskiCY{
  M.~Kruczenski, R.~Roiban, A.~Tirziu and A.~A.~Tseytlin,
  ``Strong-coupling expansion of cusp anomaly and gluon amplitudes from quantum
  open strings in $AdS_5 \times S^5$,''
  arXiv:0707.4254 [hep-th].
}

\nref\DrummondCF{
  J.~M.~Drummond, J.~Henn, G.~P.~Korchemsky and E.~Sokatchev,
  ``On planar gluon amplitudes/Wilson loops duality,''
  arXiv:0709.2368 [hep-th].
}

\Title
{\vbox{
\baselineskip12pt
\hbox{Brown-HET-1490}
}}
{\vbox{
\centerline{New Dual Conformally Invariant Off-Shell Integrals}
}}

\centerline{
Dung Nguyen, Marcus Spradlin and Anastasia Volovich
}

\vskip .5in
\centerline{Brown University}
\centerline{Providence, Rhode Island 02912 USA}

\vskip .5in
\centerline{\bf Abstract}

Evidence has recently emerged for 
a hidden symmetry of scattering amplitudes in
${\cal N} = 4$ super Yang-Mills theory called dual conformal
symmetry.
At weak coupling the presence
of this symmetry has been observed through five loops, while
at strong coupling the symmetry has been shown to have a natural
interpretation in terms of a T-dualized $AdS_5$.
In this paper we study dual conformally invariant
off-shell
four-point Feynman diagrams.
We classify all such diagrams through four loops and evaluate 10
new off-shell integrals in terms of Mellin-Barnes representations,
also finding explicit expressions for their infrared singularities.

\Date{September 2007}

\listtoc
\writetoc

\newsec{Introduction}

Recent work on
${\cal N} = 4$ super Yang-Mills theory
has unlocked rich hidden structure in planar scattering amplitudes
which indicates the exciting possibility of obtaining
exact formulas for certain amplitudes.
At weak coupling it has been observed
at two and three loops~\refs{\AnastasiouKJ,
\BernIZ} that the planar
four-particle amplitude satisfies certain iterative relations
which, if true to all loops, suggest that the full planar amplitude 
${\cal A}$ sums to the simple form
\eqn\loga{
\log({\cal A}/{\cal A}_{\rm tree}) = 
({\rm IR~divergent~terms}) + {f(\lambda) \over 8} \log^2(t/s) +
c(\lambda) + 
\cdots,
}
where $f(\lambda)$ is the cusp anomalous dimension, $s$ and $t$ are
the usual Mandelstam invariants, and the dots indicate terms which
vanish as the infrared regulator is removed.
Evidence for similar structure in the five-particle amplitude has
been found at two loops~\refs{\CachazoTJ,\BernVW}.
At strong coupling, Alday and Maldacena~\AldayHR\ have recently
given a prescription for calculating gluon scattering
amplitudes via AdS/CFT and demonstrated that the 
structure~\loga\ holds for large $\lambda$ as well.

An important role in this story is evidently played by a
somewhat mysterious symmetry of ${\cal N} = 4$ Yang-Mills
theory which has been called `dual conformal' symmetry in~\DrummondAU.
This symmetry, which is apparently unrelated to the conventional conformal
symmetry of ${\cal N} = 4$ Yang-Mills, acts as conformal transformations
on the variables $x_i \equiv k_i - k_{i+1}$, where $k_i$ are
the cyclically ordered momenta of the particles participating in a
scattering process.
It is important to emphasize that dual conformal invariance is a property
of planar amplitudes only.
Although somewhat mysterious at weak coupling,
the Alday-Maldacena prescription provides a geometrical interpretation
which makes dual conformal symmetry manifest at strong coupling.
One of the steps in their construction involves
T-dualizing along the four directions of $AdS_5$ parallel to the
boundary, and
dual conformal symmetry is the just isometry
of this T-dualized $AdS_5$.

The generalized unitarity methods
\refs{\BernZX,\BernCG,\BernSC,\BrittoNC,\BuchbinderWP}
which are used to construct the dimensionally-regulated multiloop
four-particle amplitude in ${\cal N} = 4$ Yang-Mills theory express the
ratio ${\cal A}/{\cal A}_{\rm tree}$ as a sum of certain scalar
Feynman integrals---the same kinds of integrals that would appear
in $\phi^n$ theory, but with additional scalar factors
in the numerator.
However dimensional regularization explicitly breaks dual conformal
symmetry, so the authors of~\DrummondAU\ used an alternative
regularization which consists of letting the four
external momenta in these scalar integrals
satisfy $k_i^2 = \mu^2$ instead of zero.
Then they observed that the particular Feynman integrals
which contribute to the dimensionally-regulated amplitude are precisely
those integrals which, if taken off-shell, are finite and
dual conformally
invariant in four dimensions (at least through
five loops, which is as far as the contributing integrals
are currently known~\refs{\GreenSW,\BernNH,\BernIZ,\BernEW,\BernCT}).

Off-shell dual conformally invariant integrals
have a number of properties which make them vastly simpler
to study than their on-shell cousins.
First of all they are finite in four dimensions, whereas an
$L$-loop on-shell dimensionally regulated integral has a complicated
set of infrared poles starting at order $\epsilon^{-2 L}$.
Moreover in our experience it has always proven possible to obtain
a one-term Mellin-Barnes representation for any off-shell integral, several
examples of which are shown explicitly in section 4.
In contrast, on-shell integrals typically can only be written as a sum of
many (at four loops, thousands or even tens of thousands of) separate terms
near $\epsilon = 0$.  It would be inconceivable to include a full expression
for any such integral in a paper.

Secondly, the relative simplicity of off-shell integrals is such that
a simple analytic expression for the off-shell $L$-loop 
ladder diagram was already obtained several years
ago~\refs{\UsyukinaJD,\UsyukinaCH,\BroadhurstIB}
(and generalized to arbitrary dimensions in~\IsaevTK).
For the on-shell ladder diagram
an analytic expression
is well-known at one-loop, but it
is again difficult to imagine that a simple analytic formula for any
$L$ might
even be possible.

Finally, various off-shell integrals have been observed to satisfy 
apparently highly nontrivial
relations called `magic identities' in~\DrummondRZ.
There it was proven that the three-loop ladder diagram and the three-loop
tennis court
diagram
are precisely equal to each other
in four dimensions
when taken off-shell.
Moreover a simple diagrammatic argument was given which allows one
to relate various classes of integrals to each other at any number of loops.
No trace of this structure is evident when the same integrals
are taken on-shell in $4 - 2 \epsilon$ dimensions.

Hopefully
these last few paragraphs serve to explain our enthusiasm for 
off-shell integrals.
Compared to our recent experience~\refs{\CachazoAZ,\CachazoAD}
with on-shell integrals, which required significant supercomputer time
to evaluate, we find that the off-shell 
integrals we study here are essentially trivial to evaluate.

Unfortunately however there is a very significant drawback to working
off-shell, which is that although we know (through five loops) which
scalar Feynman integrals contribute to the dimensionally-regulated
on-shell amplitude, we do not know which integrals contribute to the
off-shell amplitude.  In fact it is not even clear that one can
in general provide a meaningful definition of the `off-shell amplitude.'
Taking $k_i^2 = \mu^2$ in a scalar integral seems to be a relatively
innocuous step but we must remember that although
they are expressed in terms of scalar integrals, the amplitudes
we are interested in are really
those of non-abelian gauge bosons.  In this light relaxing
the on-shell condition $k_i^2 = 0$ does not seem so innocent.
If it is possible to consistently define a general off-shell amplitude
in ${\cal N}=4$ Yang-Mills then we would expect to see
as $\mu^2 \to 0$
the universal
leading infrared
singularity
\eqn\bbb{
\log \left({\cal A}/{\cal A}_{\rm tree}
\right) = - {f(\lambda) \over 8} \log^2(\mu^4/s t)
+ {\rm less~singular~terms},
}
where $f(\lambda)$ is the same cusp anomalous dimension 
appearing in~\loga.
If an equation of the form \bbb\ could be made to work with off-shell
integrals, it would provide a method for computing the cusp anomalous
dimension that is vastly simpler than that of reading it off
from on-shell integrals as in~\refs{\BernEW,\CachazoAZ,\CachazoAD}.

The outline of this paper is as follows.  In section 2 we review
the definition and diagrammatic properties of dual conformal integrals.
In section 3 we present the classification of off-shell dual conformal
diagrams
through four loops.
In section 4 we evaluate 10 new dual conformal integrals in terms
of Mellin-Barnes representations 
(finding two new  `magic identities') and present explicit formulas for their
behaviour in the infrared limit $\mu^2 \to 0$.

\newsec{Properties of Dual Conformal Integrals}

We begin by reviewing the definition and
diagrammatic properties of dual conformal integrals
following~\refs{\DrummondRZ,\DrummondAU}.
By way of illustration we consider first the one-loop diagram
shown in Fig.~1.
Black lines depict the underlying scalar Feynman diagram, with each internal
line associated to a $1/p^2$ propagator as usual. Each dotted red line
indicates a numerator factor $(p_1 + p_2 + \cdots + p_n)^2$
where the $p_i$ are the momenta flowing through the black lines
that it crosses.
Of course momentum conservation at each
vertex guarantees that a numerator factor only depends on where
the dotted red line begins and ends, not on the particular path
that it traverses through the diagram.

\fig{The one-loop scalar box diagram with conformal numerator
factors indicated
by the dotted red lines.}
{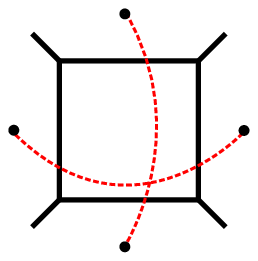}

We adopt a standard convention
(see for example~\AnastasiouKJ) that each four-dimensional
loop momentum
integral comes with a normalization factor of $1/i \pi^2$.
Thus the diagram shown in Fig.~1 corresponds to the integral
\eqn\ione{
{\cal I}^{(1)}(k_1,k_2,k_3,k_4) = \int {d^4 p_1 \over i \pi^2}
{(k_1 + k_2)^2 (k_2 + k_3)^2 \over p_1^2 (p_1 - k_1)^2 (p_1 - k_1 - k_2)^2
(p_1 + k_4)^2}.
}
We regulate this infrared divergent integral by taking
the external legs off-shell, choosing for simplicity all of the
`masses' $k_i^2 = \mu^2$ to be the same.
A different possible infrared regulator that one might consider would
be to replace the $1/p^2$ propagators by massive propagators
$1/(p^2 - m^2)$, but we
keep all internal lines strictly massless.

Following~\refs{\DrummondRZ,\DrummondAU} we then pass to dual coordinates
$x_i$ by taking
\eqn\aaa{
k_1 = x_{12}, \qquad k_2 = x_{23}, \qquad k_3 = x_{34}, \qquad
k_4 = x_{41}, \qquad p_1 = x_{15},
}
where $x_{ij} \equiv x_i - x_j$, so that~\ione\ becomes
\eqn\udy{
{\cal I}^{(1)}(x_1,x_2,x_3,x_4)
= \int {d^4 x_5 \over i \pi^2} {x_{13}^2 x_{24}^2
\over x_{15}^2 x_{25}^2 x_{35}^2 x_{45}^2}.
}
This expression is now
easily seen to be invariant under arbitrary conformal
transformations on the $x_i$.
This invariance is referred
to as `dual conformal' symmetry in~\DrummondRZ\ because it
should not be confused with the familiar conformal
symmetry of ${\cal N} = 4$ Yang-Mills.  The coordinates $x_i$ here are
momentum variables and are not simply related to the position space
variables on which the usual conformal symmetry acts.

\fig{The one-loop scalar box with dotted red lines indicating
numerator factors and thick blue lines showing the dual diagram.}{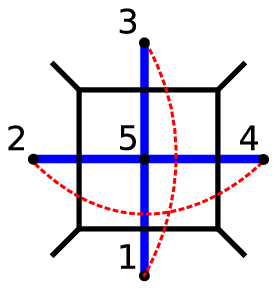}

To analyze the dual conformal invariance of a diagram
it is convenient to
consider its dual diagram\foot{The fact that nonplanar graphs do not
have duals is consistent with the observation that dual conformal symmetry
is apparently a property only of the planar limit.}.
In Fig.~2
we have labelled the vertices of the dual diagram to Fig.~1 in accord
with the expression~\udy.  The numerator factors correspond to
dotted
red lines as before, while denominator factors in the dual expression~\udy\ 
correspond to the thick blue lines in the dual diagram.
Neighboring external faces are not connected by thick blue lines because
the associated propagator is absent.

With this notation set up  it is easy to
formulate a diagrammatic rule for determining whether
a diagram can correspond to a dual conformal integral~\BernEW.
We associate to each face in the diagram (i.e., each vertex
in the dual diagram)
a weight which is equal to the number of thick blue lines attached to that
face minus the number of dotted red lines.
Then a diagram is dual conformal if
the weight of each of the four external faces is zero and
the weight of each internal face is four (to cancel the weight
of the corresponding loop integration $\int d^4 x$).
Consequently, no tadpoles, bubbles or triangles are allowed in
the Feynman diagram,
each square must be associated
with no numerator factors, each pentagon must be associated with one
numerator factor,~etc.

We distinguish slightly between dual conformal {\it diagrams}
and dual conformal {\it integrals}.
The latter are all diagrams satisfying the diagrammatic rule given above.
However in~\DrummondAU\ it was pointed out that not all dual
conformal diagrams
give rise to
integrals that are
finite off-shell in four dimensions.
Those that are not finite can only be defined with a regulator (such as
dimensional regularization) that breaks the dual conformal symmetry and
hence cannot be considered dual conformal integrals.

\fig{Two examples of three-loop dual conformal diagrams.}{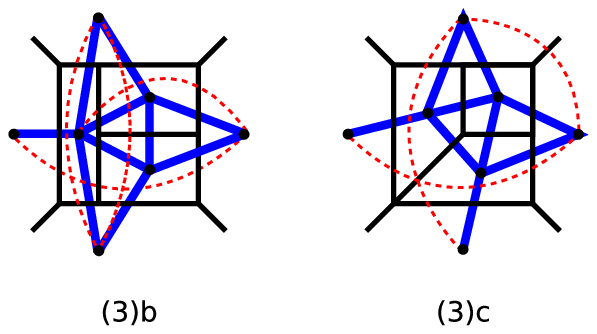}

In Fig.~3 we show two diagrams that are easily seen to be dual conformal
according to the above rules.
The diagram on the left is the well-known three-loop tennis court
${\cal I}^{(3)b}$~\BernIZ.  The diagram on the right
demonstrates a new feature
that is possible only when the external lines are taken off-shell.
The dotted red line connecting the top external face to the
external face on the right
crosses only one external line and is therefore associated with
a numerator factor of $k_i^2 = \mu^2$.
Such an integral is absent when we work on-shell.  Of course
{\it absent} does not necessarily mean that an integral {\it vanishes}
if we first calculate it for finite $\mu^2$ and then take $\mu^2 \to 0$.
Indeed we will see below that ${\cal I}^{(3)c}
\sim \ln^3(\mu^2)$ in the infrared limit.

An important and well-known feature of
four-point dual conformal integrals
is that they are constrained by the symmetry
to be a function only of the conformally invariant
cross-ratios
\eqn\aaa{
u = {x_{12}^2 x_{34}^2 \over x_{13}^2 x_{24}^2}, \qquad
v = {x_{14}^2 x_{23}^2 \over x_{13}^2 x_{24}^2}.
}
Since we have chosen to take all external
momenta to have the same value of $k_i^2 = \mu^2$, we see that $u$ and
$v$ are are actually both equal to
\eqn\aaa{
x \equiv {\mu^4 \over s t},
}
where $s = (p_1 + p_2)^2$ and $t= (p_2 + p_3)^2$ are the usual
Mandelstam invariants.
Therefore we can express any dual conformal integral as a function
of the single variable $x$.
One immediate consequence of this observation is that any dual conformal
integral is invariant under rotations and reflections of the corresponding
diagram, since $x$ itself is invariant under such permutations.
A second consequence is that any degenerate integral (by which we mean one
where two or more of the external momenta enter the diagram at the same
vertex) must evaluate to a constant, independent of $x$.

\newsec{Classification of Dual Conformal Diagrams}

\subsec{Algorithm}

Let us now explain a systematic algorithm to enumerate all possible
off-shell
dual conformal diagrams.
We use the graph generating program qgraf~\qgraf\ to generate all 
scalar
1PI\foot{We do not know of a general proof that no one-particle reducible
diagram can be dual conformal but it is easy to check through four loops
that there are no such examples.}
four-point topologies with no tadpoles or bubbles, and throw away any
that are non-planar or contain triangles since these cannot be made
dual conformal.  After these cuts there remain $(1,1,4,25)$ distinct
topologies at $(1,2,3,4)$ loops respectively
respectively.  We adopt the naming conventions from 
reference \BernEW\ which displays
24 out of the
25 four-loop topologies,
omitting the one we call $h$ in Fig.~9 since it vanishes
on-shell
in dimensional regularization.  Note also that the topologies
$e_5$ and $c_1$ shown there are actually the same.

The next step is to try adding numerator factors
to render each diagram dual conformal.  Through three loops
this is possible in a unique way for each topology, but
at four loops there are three topologies (shown in Fig.~4) that
cannot be made dual conformal at all while six topologies
($b_1$, $c$, $d$, $e$, $e_2$ and $f$, shown below) each admit two
distinct choices of numerator factors making the diagram dual conformal.

\fig{These are the three planar four-point 1PI four-loop tadpole-,
bubble- and triangle-free topologies
that cannot be made into dual conformal diagrams by the addition
of any numerator factors.
In each case the obstruction is that there is a single
pentagon whose excess weight cannot be cancelled by any numerator
factor because the pentagon borders on all of the external faces.
 (There are no examples of this below
four loops.)}{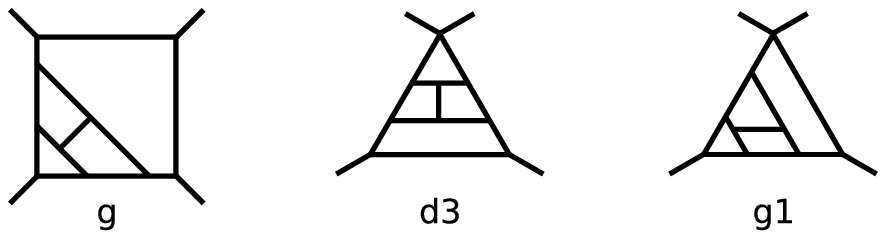}

We remark that we exclude from our classification
certain `trivial' diagrams that
can be related to others by rearranging numerator
factors connected only to external faces.  For example, consider
the two diagrams shown in Fig.~5.  Clearly both are
dual conformal, but they differ from each other only by an overall factor
of $x = \mu^4/s t$.
In cases such as this we include in our classification only the diagram
with the fewest number of $\mu^2$ powers in the numerator, thereby
choosing the integral that is most singular in the $\mu^2 \to 0$ limit.
In the example of Fig.~5 we therefore exclude the diagram on the right,
which actually vanishes in the $\mu^2 \to 0$ limit, in favor of the diagram
on the left, which behaves like $\ln^2 (\mu^2)$.

\fig{Two dual conformal diagrams that differ only by an overall factor.
As explained in the text we resolve such ambiguities by choosing the integral
that is most singular in the $\mu^2 \to 0$ limit, in this example
eliminating the diagram on the right.}
{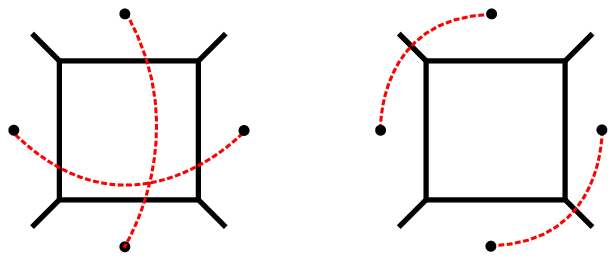}

Another possible algorithm, which we have used to double check our
results, is to first use the results of~\KajantieHV\ to generate all planar
1PI vacuum graphs and then enumerate all possible ways of attaching
four external legs so that no triangles or bubbles remain.

\subsec{Results}

Applying the algorithm just described,
we find a total of $(1,1,4,28)$ distinct
dual conformal
diagrams respectively at $(1,2,3,4)$ loops.
While $(1,1,2,10)$ of these diagrams have appeared
previously in the literature on dual conformal
integrals~\refs{\DrummondRZ,\BernEW,\BernCT},
the remaining $(0,0,2,18)$ that only exist off-shell are new to this
paper.
We classify all of these diagrams into four groups,
according to whether or not they are finite in four dimensions,
and according to whether or not the numerator contains any explicit
factors of $\mu^2$.  Hence we define:

Type I diagrams are finite in four dimensions and have
no $\mu^2$ factors.

Type II diagrams are divergent in four dimensions and have no $\mu^2$
factors.

Type III diagrams are finite in four dimensions and have $\mu^2$ factors.

Type IV diagrams are divergent in four dimensions and have $\mu^2$ factors.

Type I and II diagrams through five loops have been classified,
and some of their properties studied,
in~\refs{\DrummondRZ,\BernEW,\BernCT,\DrummondAU}.
In particular, it has been observed in these references that it is
precisely the type I integrals that contribute to the
dimensionally regulated on-shell four-particle
amplitude (at least through five loops).
We display these diagrams through four loops
in Figs.~6 and~7.
The new type III and IV diagrams that
only exist off-shell are shown respectively in Figs.~8 and~9.
Each diagram is given a name of the form ${\cal I}^{(L)i}$ where
$L$ denotes the number of loops and $i$ is a label.
The one- and two-loop diagrams are unique and do not require a label.
Below we will also use ${\cal I}^{(L)}$ to refer to the $L$-loop
ladder diagram (specifically, ${\cal I}^{(1)}$,
${\cal I}^{(2)}$, ${\cal I}^{(3)a}$ and ${\cal I}^{(4)a}$ for
$L=1,2,3,4$).

We summarize the results of our classification in the following
table showing the number of dual conformal diagrams
of each type at each loop order:
\eqn\aaa{
\matrix{
L &\vrule height2.75ex depth1.25ex&
{\rm I} & {\rm II} & {\rm III} & {\rm IV} \cr
\noalign{\hrule height 0.6pt} \cr
{\rm 1} &\vrule height2.75ex& 1 & 0 & 0 & 0 \cr
{\rm 2} &\vrule height2.75ex& 1 & 0 & 0 & 0 \cr
{\rm 3} &\vrule height2.75ex& 2 & 0 & 2 & 0 \cr
{\rm 4} &\vrule height2.75ex& 8 & 2 & 9 & 9
}
}

\fig{Type I:  Here we show all dual conformal diagrams through four loops that
are finite off-shell in four dimensions and
have no explicit numerator factors of $\mu^2$.
These are precisely the integrals which contribute to the
dimensionally-regulated on-shell four-particle
amplitude~\refs{\GreenSW,\BernNH,\BernIZ,\BernEW}.
For clarity we suppress an overall factor of $st$ in each diagram.
}{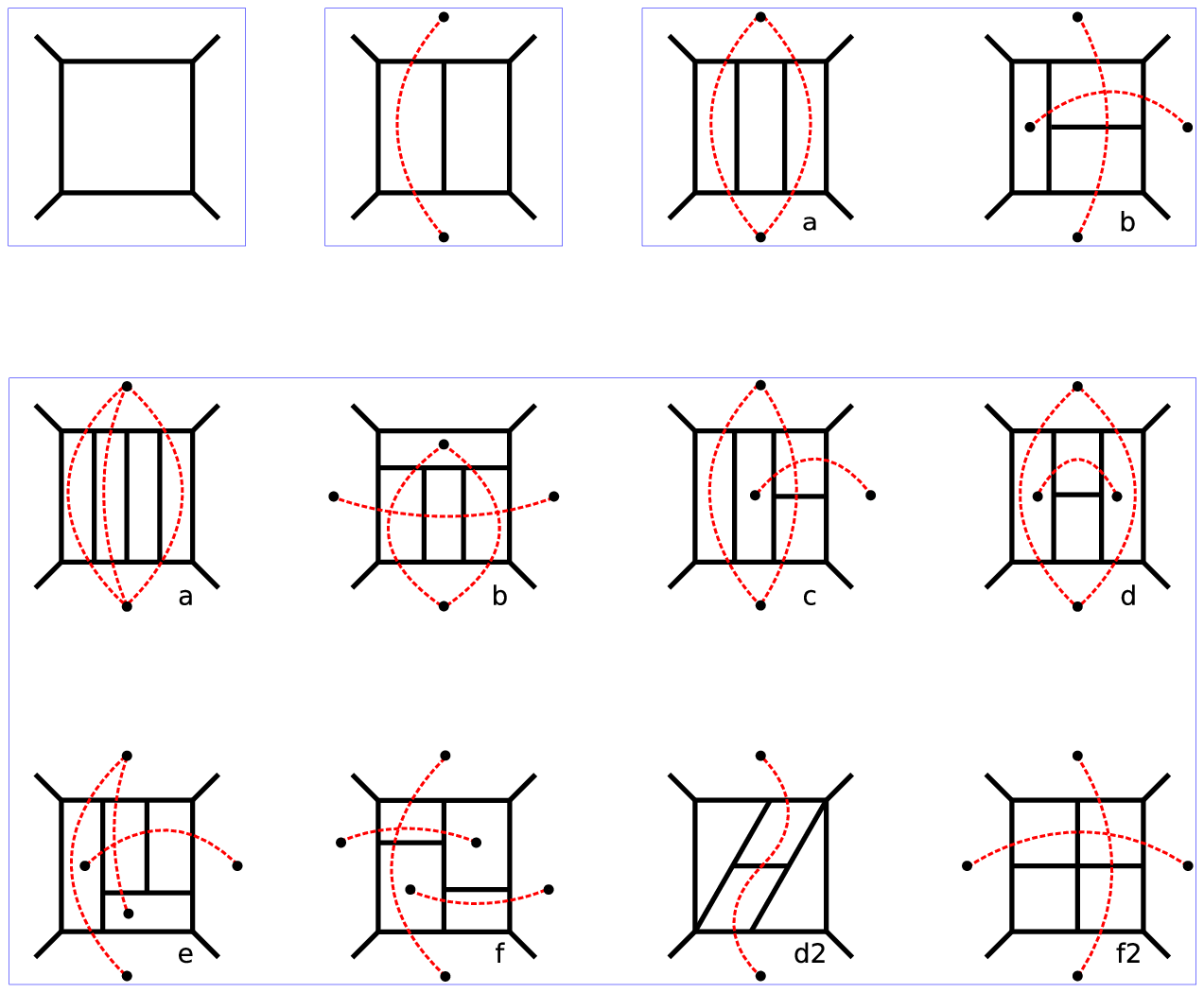}

\fig{Type II:  These two diagrams
have no explicit factors of $\mu^2$ in the numerator and
satisfy the diagrammatic criteria
of dual conformality, but the corresponding off-shell integrals diverge
in four dimensions~\DrummondAU.
(There are no examples of this type below four loops).}{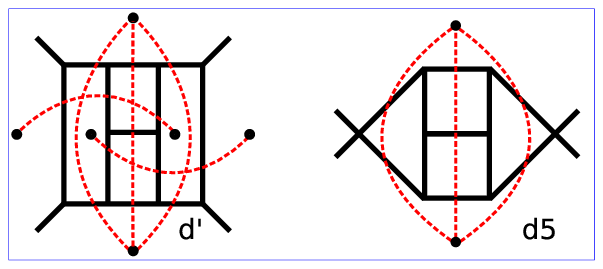}

\fig{Type III:
Here we show all diagrams through four loops that correspond
to dual conformal integrals in four dimensions with explicit
numerator factors of $\mu^2$.  (There are no examples of this
type below three loops).
In the bottom row we have isolated three degenerate diagrams
which are constrained by dual conformal invariance
to be equal to pure numbers (independent of $s$, $t$ and $\mu^2$).
}{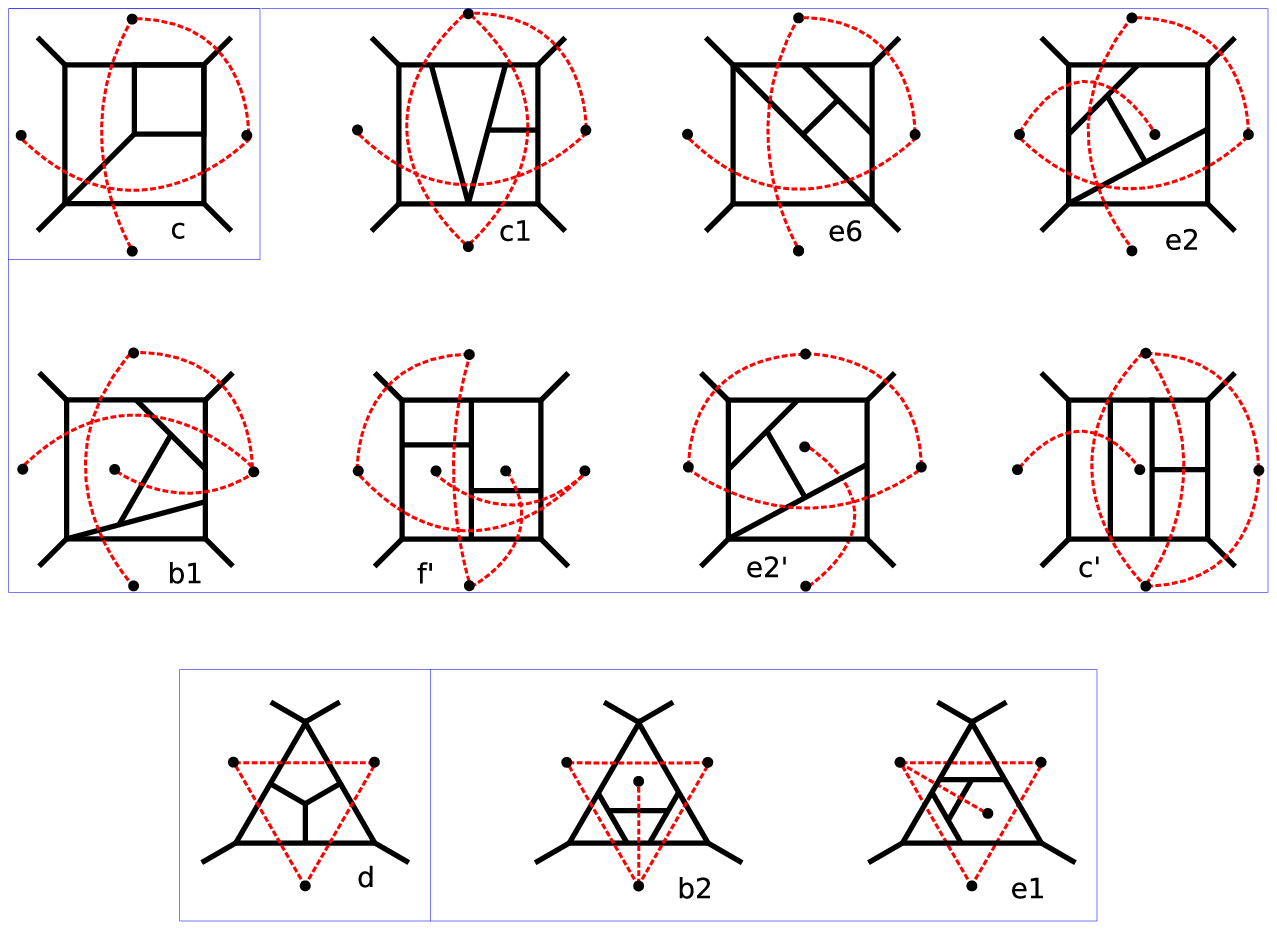}

\fig{Type IV: All remaining dual conformal diagrams.
All of the corresponding off-shell integrals diverge in four dimensions.
}{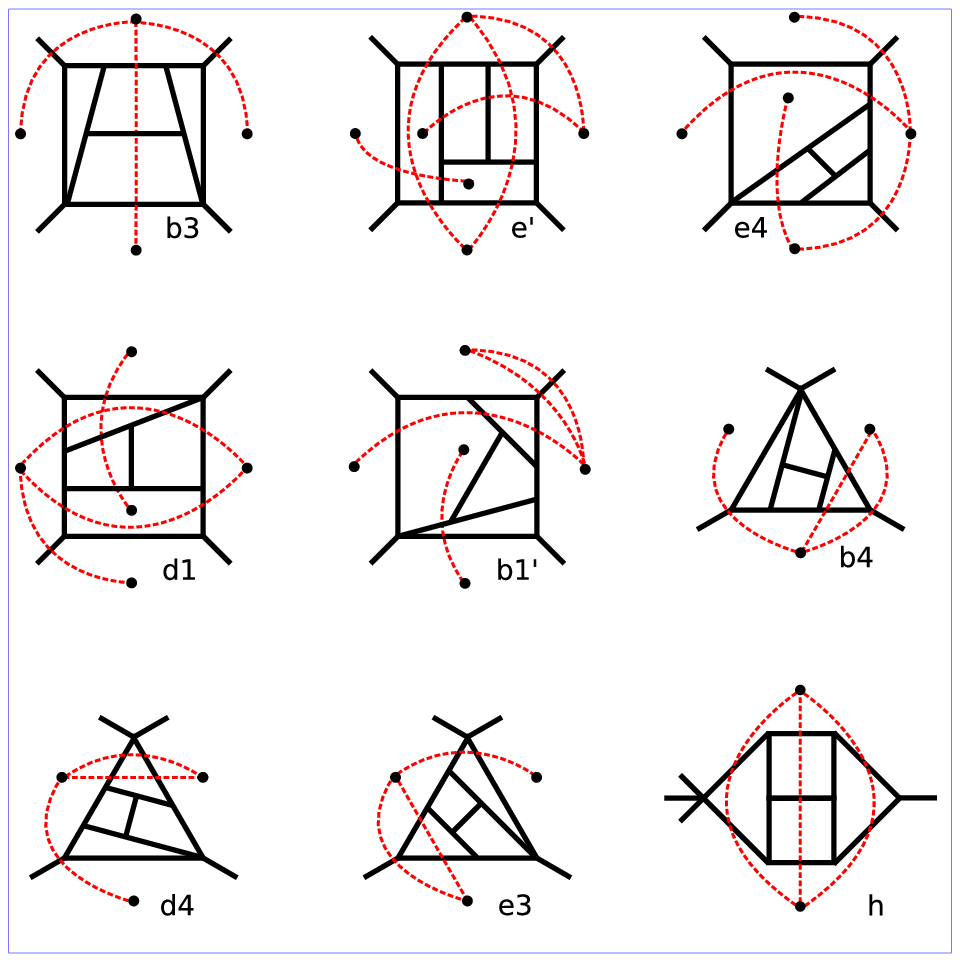}

\newsec{Evaluation of Dual Conformal Integrals}

In this section we describe the evaluation of dual conformal integrals.
Even though dual conformal integrals are finite in four dimensions
we evaluate them
by first dimensionally regulating the integral
to $D = 4 - 2 \epsilon$ dimensions, and then analytically continuing
$\epsilon$ to zero using
for example algorithms described in~\refs{\SmirnovGC,\TauskVH,\CzakonRK}.
For a true dual conformal invariant integral the result of this
analytic continuation will be an integral that is finite in four dimensions,
so that we can then freely set $\epsilon = 0$.
However the type II and type IV diagrams shown in Figs.~6 and~8 turn
out to not be finite in four dimensions (as can be verified either
by direct calculation, or by applying the argument used in~\DrummondAU\ to
identify divergences).
This leaves (1,1,4,17) integrals to be evaluated at (1,2,3,4) loops
respectively.

\subsec{Previously known integrals}

Here we briefly review the
(1,1,2,5) off-shell dual conformal integrals that
have already been evaluated in the literature.

The first class of integrals that have been evaluated off-shell
are those corresponding to the ladder diagrams ${\cal I}^{(1)}$,
${\cal I}^{(2)}$, ${\cal I}^{(3)a} \equiv {\cal I}^{(3)}$
and ${\cal I}^{(4)a}\equiv {\cal I}^{(4)}$.
In fact an explicit formula for the off-shell $L$-loop ladder diagram
was given in the remarkable paper~\UsyukinaCH.
The function $\Phi^{(L)}(x,x)$ in that paper corresponds precisely
to our conventions in defining the ladder diagrams ${\cal I}^{(L)}$
(including the appropriate conformal numerator factors), so we copy
here their result
\eqn\amazing{
{\cal I}^{(L)}(x) =
{2 \over \sqrt{1 - 4 x}}
\left[
{(2 L)! \over L!^2} \Li_{2 L}(- y)
+ \sum_{{k,l=0} \atop {k + l~{\rm even}}}^L {(k + l)! (1 - 2^{1 - k - l})
\over k! l! (L - k)! (L - l)!} \zeta(k + l) \log^{2 L - l - k}y
\right],
}
where
\eqn\aaa{
y = {2 x \over 1 - 2 x + \sqrt{1 - 4 x}}.
}

The second class of integrals that have been evaluated off-shell
are those that can be proven equal to ${\cal I}^{(L)}$ using the
`magic identities' of~\DrummondRZ.  There it was shown that
\eqn\mone{
{\cal I}^{(3)a} = {\cal I}^{(3)b} = {\cal I}^{(3)}
}
and
\eqn\mtwo{
{\cal I}^{(4)a} = {\cal I}^{(4)b} = {\cal I}^{(4)c} =
{\cal I}^{(4)d} = {\cal I}^{(4)e} = {\cal I}^{(4)}.
}
These identities appear to be highly nontrivial 
and are only valid
for the off-shell integrals in four dimensions;
certainly no hint of any relation between these integrals
is apparent when they
are taken on-shell and evaluated
in $4 - 2 \epsilon$ dimensions as in~\refs{\BernIZ,\BernEW,\CachazoAD}.
Moreover in~\DrummondRZ\ the relations~\mone\ and~\mtwo\ were given
a simple diagrammatic interpretation which can be utilized to
systematically identify equalities between
certain integrals at any number of loops.

\subsec{New integrals}

As indicated above we evaluate these integrals by starting with
Mellin-Barnes representations in $4 - 2 \epsilon$ dimensions
and then analytically continuing $\epsilon$ to zero.
A very useful Mathematica code which automates this process
has been written by Czakon~\CzakonRK.  We found however
that this implementation was too slow to handle some of the
new integrals in a reasonable amount of time so we implemented
the algorithm in a C program instead.  The most difficult
off-shell
integral we have evaluated,
${\cal I}^{(4)f2}$, starts off in $4 - 2 \epsilon$ dimensions
as a 24-fold Mellin-Barnes representation (far more complicated
than any of the on-shell four-loop integrals considered
in~\refs{\BernEW,\CachazoAZ}, which require at most 14-fold
representations), yet the analytic continuation to 4 dimensions
takes only a fraction of a second in C.
In what follows we display Mellin-Barnes representations
for the various integrals in 4 dimensions, after the analytic
continuation has been performed and $\epsilon$ has been set to zero.

The most surprising aspect of the formulas given below is that
we are able to write each integral in terms of just a single
Mellin-Barnes integral in four dimensions.  This stands
in stark contrast to dimensionally regulated on-shell integrals,
for which the analytic continuation towards $\epsilon = 0$ can
generate (at four loops, for example) thousands or even tens
of thousands of terms.  For the off-shell integrals studied
here something rather amazing happens:  the analytic continuation
still produces thousands of terms (or more), but for each off-shell
integral
it turns out that only one of the resulting terms is non-vanishing
at $\epsilon = 0$, leaving in each case only a single
Mellin-Barnes integral in four dimensions.

This surprising result
is not automatic but
depends on a number of factors, including the
choice of initial Mellin-Barnes representation,
the choice
of integration contour for the Mellin-Barnes variables $z_i$,
and some details of the how the analytic continuation is carried out.
All of these steps
involve highly non-unique choices, and by making different choices
it is easy to end up with more than one term that is finite in
four dimensions.
However
in such cases it is always possible to 
`reassemble' the finite terms 
into the one-term
representations shown here by shifting the contours of the
remaining integration variables.

For the new 3-loop integrals, both shown in Fig.~8, we
find the Mellin-Barnes representations\foot{All formulas in this
section are valid when the integration contours for the $z_i$ are
chosen to be straight lines parallel to the imaginary axis and such
that the arguments of all $\Gamma$ functions in the numerator
of the integrand
have
positive real part.}
\eqn\aaa{\eqalign{
{\cal I}^{(3)c}
&= - \int {d^5 z \over (2 \pi i)^5} \, x^{z_2}\,
\Gamma (-{z_1}) \Gamma (-{z_2}) \Gamma ({z_2}+1) \Gamma
    (-{z_2}+{z_3}+1) \Gamma ({z_1}+{z_3}-{z_4}+1) \cr
& \qquad \Gamma
    (-{z_4})^2 \Gamma ({z_4}-{z_3}) \Gamma
    (-{z_2}-{z_5})^2 \Gamma ({z_1}-{z_4}-{z_5}+1) \cr
& \qquad \Gamma
    ({z_1}-{z_2}+{z_3}-{z_4}-{z_5}+1) \Gamma
    (-{z_1}+{z_4}+{z_5}) \Gamma ({z_2}+{z_4}+{z_5}+1) \cr
& \qquad \Gamma (-{z_1}+{z_2}+{z_4}+{z_5}) \Gamma
    (-{z_1}-{z_3}+{z_4}+{z_5}-1)/(
\Gamma (1-{z_4}) \cr
& \qquad \Gamma (-{z_2}-{z_5}+1) \Gamma
    ({z_1}-{z_2}+{z_3}-{z_4}-{z_5}+2) \Gamma
    (-{z_1}-{z_2}+{z_4}+{z_5}) \cr
& \qquad \Gamma
    (-{z_1}+{z_2}+{z_4}+{z_5}+1))
}}
and
\eqn\threed{\eqalign{
{\cal I}^{(3)d}
&= \int {d^4 z \over (2 \pi i)^z}\,
\Gamma (-{z_1}) \Gamma ({z_1}+1) \Gamma (-{z_2}) \Gamma
    ({z_1}+{z_2}-{z_3}+1) \Gamma (-{z_3})^2 \Gamma
    ({z_3}-{z_1}) \cr
& \qquad \Gamma ({z_2}-{z_3}-{z_4}+1) \Gamma
    ({z_1}+{z_2}-{z_3}-{z_4}+1) \Gamma (-{z_4})^2 \Gamma
    ({z_3}+{z_4}+1) \cr
& \qquad \Gamma (-{z_2}+{z_3}+{z_4}) \Gamma
    (-{z_1}-{z_2}+{z_3}+{z_4})/(
\Gamma (1-{z_1}) \Gamma (1-{z_3}) \Gamma (1-{z_4}) \cr
& \qquad \Gamma
    ({z_1}+{z_2}-{z_3}-{z_4}+2) \Gamma
    (-{z_2}+{z_3}+{z_4}+1)).
}}
As expected, ${\cal I}^{(3)d}$ is $x$-independent because the corresponding
diagram is degenerate.  Upon 
evaluating~\threed\ numerically (using CUBA~\HahnFE) we find
\eqn\threedval{
{\cal I}^{(3)d} \approx 20.73855510
}
with a reported
estimated numerical uncertainty smaller than the last digit shown.

Finally we have 12 off-shell four-loop integrals left to evaluate,
corresponding
to the 9 diagrams shown in Fig.~8, along with three of the diagrams
(${\cal I}^{(4)f}$, ${\cal I}^{(4)d2}$
and ${\cal I}^{(4)f2}$) from Fig.~6.
It turns out that 4 of these 12 integrals
(${\cal I}^{(4)f}$, ${\cal I}^{(4)f'}$, ${\cal I}^{(4) e2'}$ and
${\cal I}^{(4)c'}$)
are significantly more difficult than the rest because they
apparently require analytic continuation not only in $\epsilon$ but
also in a second parameter $\nu$ parameterizing the power of the numerator
factors.  (That is, the integrals initially converge only for
$\nu < 1$ and must be analytically continued to $\nu = 1$.)
We postpone the study of these more complicated integrals to future work.

In analyzing the remaining 8 off-shell 
four-loop integrals we have found
two new `magic identities',
\eqn\magic{
{\cal I}^{(4) e2} = {\cal I}^{(4) b1}, \qquad
{\cal I}^{(4) c1} = - {\cal I}^{(4) d2}.
}
We established these results directly by deriving
Mellin-Barnes representations for these integrals and showing that
they can be related to each other under a suitable change of integration
variables.
It would certainly be interesting to understand the origin of
the relations \magic\ and to see whether the insight gained
thereby can be used to relate various dual conformal
integrals at higher loops to each other.

Mellin-Barnes representations for the 8 off-shell four-loop
integrals are:
\eqn\aaa{\eqalign{
{\cal I}^{(4)c1} &= - {\cal I}^{(4)d2}
= - \int {d^7 z \over (2 \pi i)^7} \, x^{z_2}\,
\Gamma (-{z_1}-1) \Gamma ({z_1}+2) \Gamma (-{z_2}) \Gamma
    ({z_2}+1) \cr
& \qquad \Gamma (-{z_1}-{z_3}-1) \Gamma (-{z_3}) \Gamma
    ({z_3}-{z_2}) \Gamma ({z_1}-{z_2}+{z_3}+1) \Gamma
    (-{z_4}) \cr
& \qquad \Gamma ({z_1}-{z_2}+{z_3}+{z_5}+2) \Gamma
    (-{z_6})^2
    \Gamma ({z_4}+{z_5}-{z_7}+1) \cr
& \qquad \Gamma
    (-{z_1}+{z_2}+{z_4}-{z_6}-{z_7}) \Gamma
    ({z_4}+{z_5}-{z_6}-{z_7}+1)
    \Gamma (-{z_7})^2 \Gamma
    ({z_7}-{z_5}) \cr
& \qquad \Gamma ({z_6}+{z_7}+1) \Gamma
    (-{z_4}+{z_6}+{z_7})
    \Gamma (-{z_3}-{z_4}-{z_5}+{z_6}+{z_7}-1)/ \cr
& \qquad (\Gamma (-{z_1}-{z_2}-1) \Gamma (1-{z_3}) \Gamma
    ({z_1}-{z_2}+{z_3}+2) \Gamma (1-{z_6}) \Gamma
    (1-{z_7}) \cr
& \qquad \Gamma ({z_4}+{z_5}-{z_6}-{z_7}+2) \Gamma
    (-{z_4}+{z_6}+{z_7}+1))
}}

\eqn\aaa{\eqalign{
{\cal I}^{(4)f2}
&= \int {d^{10} z \over (2 \pi i)^{10}}\, x^{z_1} \,
   \Gamma (-{z_1}-{z_{10}})^2 \Gamma (-{z_2}) \Gamma (-{z_3})
    \Gamma ({z_1}+{z_{10}}+{z_3}+1)^2 \Gamma (-{z_4}) \cr
& \qquad \Gamma (-{z_5}) \Gamma (-{z_6}) \Gamma ({z_{10}}+{z_2}+{z_6})
    \Gamma (-{z_7})^2 \Gamma ({z_4}+{z_7}+1)^2 \cr
& \qquad \Gamma
    (-{z_1}-{z_4}-{z_5}-{z_8}-2)
    (-{z_1}-{z_{10}}-{z_2}-{z_4}-{z_6}-{z_7}-{z_8}-2
    ) \cr
& \qquad \Gamma (-{z_8}) \Gamma ({z_1}+{z_8}+2)
    ({z_2}+{z_4}+{z_5}+{z_7}+{z_8}+2) \cr
& \qquad \Gamma
    (-{z_2}-{z_3}-{z_4}-{z_5}-{z_9}-2) \Gamma
    (-{z_{10}}-{z_2}-{z_3}-{z_6}-{z_9}) \Gamma (-{z_9}) \cr
& \qquad \Gamma ({z_1}+{z_{10}}+{z_2}+{z_3}+{z_6}+{z_9}+2) \cr
& \qquad \Gamma
    ({z_{10}}+{z_2}+{z_3}+{z_4}+{z_5}+{z_6}+{z_9}+2)
    \Gamma (-{z_1}+{z_3}-{z_8}+{z_9}) \cr
& \qquad \Gamma
    ({z_2}+{z_4}+{z_5}+{z_6}+{z_8}+{z_9}+2)/(
\Gamma ({z_1}+{z_{10}}+{z_3}+2) \Gamma (-{z_4}-{z_5}) \cr
& \qquad \Gamma (-{z_1}-{z_{10}}-{z_7}) \Gamma ({z_4}+{z_7}+2)
    \Gamma (-{z_1}-{z_{10}}-{z_2}-{z_6}-{z_8}) \cr
& \qquad \Gamma
    (-{z_3}-{z_9}) \Gamma
    (-{z_1}+{z_{10}}+{z_2}+{z_3}+{z_6}-{z_8}+{z_9}) \cr
& \qquad \Gamma
    ({z_1}+{z_{10}}+{z_2}+{z_3}+{z_4}+{z_5}+{z_6}+
    {z_8}+{z_9}+4))
}}

\eqn\aaa{\eqalign{
{\cal I}^{(4)e6}
&= - \int {d^5 z \over (2 \pi i)^5}\, x^{z_2}\,
\Gamma (-{z_1}) \Gamma (-{z_2})^4 \Gamma ({z_2}+1)^2 \Gamma
    (-{z_3}) \Gamma ({z_3}+1) \cr
& \qquad \Gamma ({z_1}+{z_3}-{z_4}+1) \Gamma (-{z_4})^2
    \Gamma ({z_4}-{z_3}) \Gamma
    ({z_1}-{z_4}-{z_5}+1) \cr
& \qquad \Gamma
    ({z_1}+{z_3}-{z_4}-{z_5}+1) \Gamma (-{z_5})^2 \Gamma
    ({z_4}+{z_5}+1) \Gamma (-{z_1}+{z_4}+{z_5}) \cr
& \qquad \Gamma
    (-{z_1}-{z_3}+{z_4}+{z_5})/(
\Gamma (-2 {z_2}) \Gamma
    (1-{z_3}) \Gamma (1-{z_4}) \Gamma
    (1-{z_5}) \cr
& \qquad \Gamma ({z_1}+{z_3}-{z_4}-{z_5}+2) \Gamma
    (-{z_1}+{z_4}+{z_5}+1))
}}

\eqn\aaa{\eqalign{
{\cal I}^{(4)e2} &= {\cal I}^{(4)b1}
= - \int {d^7 z \over (2 \pi i)^7}\, x^{z_3}\,
\Gamma (-{z_1}) \Gamma (-{z_3}) \Gamma ({z_3}+1) \Gamma
    ({z_3}-{z_2}) \Gamma (-{z_4})^2 \cr
& \qquad \Gamma ({z_2}-{z_3}+{z_4}+1)^2\Gamma (-{z_5})^2 \Gamma
    ({z_1}+{z_2}-{z_5}-{z_6}+2) \Gamma (-{z_6})^2 \cr
& \qquad \Gamma
    ({z_5}+{z_6}+1) \Gamma
    (-{z_1}-{z_2}+{z_5}+{z_6}-1) \Gamma
    (-{z_1}-{z_2}+{z_3}+{z_5}+{z_6}-1) \cr
& \qquad \Gamma
    (-{z_4}+{z_5}-{z_7}) \Gamma
    (-{z_1}-{z_2}-{z_4}+{z_5}+{z_6}-{z_7}-1) \Gamma
    (-{z_7}) \cr
& \qquad \Gamma (-{z_3}+{z_4}+{z_7}) \Gamma
    ({z_1}+{z_2}-{z_3}+{z_4}-{z_5}+{z_7}+2) \cr
& \qquad \Gamma
    ({z_1}-{z_5}-{z_6}+{z_7}+1)/(
\Gamma ({z_2}-{z_3}+{z_4}-{z_5}+2) \Gamma
    (-{z_4}-{z_6}+1) \cr
& \qquad \Gamma
    (-{z_1}-{z_2}-{z_3}+{z_5}+{z_6}-1) \Gamma
    (-{z_1}-{z_2}+{z_3}+{z_5}+{z_6}) \cr
& \qquad \Gamma
    (-{z_4}-{z_7}+1) \Gamma
    ({z_1}+{z_2}-{z_3}+{z_4}-{z_5}-{z_6}+{z_7}+2))
}}

\eqn\aaa{\eqalign{
{\cal I}^{(4)b2}
&= {\cal I}^{(4)e1}
= \int {d^6 z \over (2 \pi i)^6}\,
\Gamma (-{z_1}) \Gamma (-{z_2}) \Gamma (-{z_3}) \Gamma
    ({z_3}+1) \Gamma (-{z_1}-{z_2}-{z_4}-2) \cr
& \qquad \Gamma
    (-{z_1}-{z_2}-{z_3}-{z_4}-2) \Gamma (-{z_4}) \cr
& \qquad \Gamma
    ({z_1}+{z_2}+{z_4}+3) \Gamma
    ({z_1}+{z_2}+{z_3}+{z_4}+3) \Gamma
    ({z_1}+{z_3}-{z_5}+1) \Gamma (-{z_5}) \cr
& \qquad \Gamma
    ({z_5}-{z_3}) \Gamma ({z_2}+{z_4}+{z_5}+2) \Gamma
    (-{z_4}-{z_5}-{z_6}-1) \Gamma (-{z_6})^2 \cr
& \qquad \Gamma
    ({z_4}+{z_6}+1)^2 \Gamma
    ({z_2}+{z_4}+{z_5}+{z_6}+2)/(
\Gamma (1-{z_3}) \cr
& \qquad \Gamma (-{z_1}-{z_2}-{z_4}-1) \Gamma
    ({z_1}+{z_2}+{z_3}+{z_4}+4) \Gamma (1-{z_5}) \cr
& \qquad \Gamma
    ({z_2}+{z_4}+{z_5}+3) \Gamma (1-{z_6}) \Gamma
    ({z_4}+{z_6}+2))
}}
We do not consider the equality 
of the two degenerate integrals ${\cal I}^{(4)b2} = {\cal I}^{(4)e1}$
to be a `magic' identity since it is easily seen to be a trivial
consequence of dual conformal invariance.
Evaluating them numerically we find
\eqn\aaa{
{\cal I}^{(4)b2} = {\cal I}^{(4)e1} = 70.59,
}
again with a reported
estimated numerical uncertainty smaller than the last digit shown.

It would certainly be interesting to obtain fully explicit analytic
results for these new integrals.  Although
this might
seem to be a formidable challenge, the fact that it has been possible
for the ladder diagrams~\amazing\ suggests that there is hope.

\subsec{Infrared singularity structure}

Finally, it is clearly of interest to isolate the infrared singularities
of the various integrals.
For the previously known integrals reviewed in subsection 3.1 we
expand~\amazing\ for small $x$, finding
\eqn\aaa{\eqalign{
{\cal I}^{(1)} &=
\log^2 x
+ {\cal O}(1)\cr
{\cal I}^{(2)} &=
{1 \over 4} \log^4 x
+ {\pi^2 \over 2} \log^2 x
+ {\cal O}(1),\cr
{\cal I}^{(3)} &=
{1 \over 36} \log^6 x
+ {5 \pi^2 \over 36} \log^4 x
+ {7 \pi^4 \over 36} \log^2 x
+ {\cal O}(1),\cr
{\cal I}^{(4)} &=
{1 \over 576} \log^8 x
+ {7 \pi^2 \over 432} \log^6 x
+ {49 \pi^4 \over 864} \log^4 x
+ {31 \pi^6 \over 432} \log^2 x
+ {\cal O}(1).
}}
For the new integrals evaluated in this paper we obtain
the small $x$ expansion directly from the Mellin-Barnes representations
given in section 3.2 by writing each one in the form
\eqn\aaa{
\int{dy \over 2 \pi i} x^y F(y),
}
shifting the $y$ contour of integration to the left until it sits
directly on the imaginary axis (picking up terms along the way from
any poles crossed), expanding the resulting integrand around $y=0$ and then
using the fact that the coefficient of the $1/y^k$ 
singularity at $y=0$
corresponds in $x$ space to the coefficient of the ${(-1)^k \over k!}
\log^k x$ singularity at $x=0$.
In this manner we find
\eqn\aaa{
{\cal I}^{(3)c} = {\zeta(3) \over 3} \log^3 x - {\pi^4 \over 30} \log^2 x
+ 14.32388625\,\log x + {\cal O}(1)
}
at three loops
and
\eqn\aaa{\eqalign{
{\cal I}^{(4)d2} = - {\cal I}^{(4)c1} &= -  {\zeta(3) \over 12} \log^5 x
+ {7 \pi^4 \over 720} \log^4 x
-6.75193310\,\log^3 x
\cr
&\qquad\qquad
+ 15.45727322\,\log^2 x
-41.26913\,\log x+ {\cal O}(1),\cr
{\cal I}^{(4)f2} &= {1 \over 144} \log^8 x + {7 \pi^2 \over 108}
\log^6 x + {149 \pi^4 \over 1080} \log^4 x\cr
&\qquad\qquad+ 64.34694867\,\log^2 x +
{\cal O}(1),\cr
{\cal I}^{(4)e6} &= -20.73855510\,\log^2 x + {\cal O}(1),\cr
{\cal I}^{(4)e2} = {\cal I}^{(4) b1} &= - {\pi^4 \over 720} \log^4 x 
+ 1.72821293\,\log^3 x
\cr
&\qquad\qquad
 - 12.84395616\,\log^2 x
 + 52.34900\,\log x
+ {\cal O}(1)
}}
at four loops,
where some coefficients have only been evaluated numerically
with an estimated uncertainly smaller than the last digit shown.
Interestingly, the coefficient of $\log^2 x$ in ${\cal I}^{(4) e6}$
appears to be precisely (minus) the value of ${\cal I}^{(3)d}$ shown
in~\threedval.  Perhaps this can be traced to the
diagrammatic relation that is
evident in Fig.~8: ${\cal I}^{(3)d}$ appears
in the `upper diagonal' of ${\cal I}^{(4)e6}$.

\newsec{Summary}

We have classified all four-point dual conformal Feynman diagrams
through four loops.  In addition to the previously known
$(1,1,2,8)$ integrals (Fig.~6) that contribute to the
dimensionally-regulated on-shell amplitude respectively at $(1,2,3,4)$ loops,
we find $(0,0,2,9)$ new dual conformal integrals
(Fig.~8)
that vanish on-shell
in $D = 4 - 2 \epsilon$ but not off-shell in $D = 4$.
There are also $(0,0,0,11)$ dual conformal diagrams
(Figs.~7 and~9)
that diverge in four dimensions even when taken off-shell and therefore
do not give rise to true dual conformal integrals.

Next we addressed the problem of evaluating
new off-shell integrals in four dimensions.
Of the
total number $(1,1,4,17)$ of such integrals, explicit results for
$(1,1,2,5)$ have
appeared previously in~\refs{\UsyukinaCH,\DrummondRZ}.
We find Mellin-Barnes representations for 
an additional $(0,0,2,8)$ integrals,
including two pairs related by new `magic identities',
and evaluate their infrared singularity
structure explicitly.
Evaluation of the
remaining $(0,0,0,4)$ integrals
${\cal I}^{(4)f}$,
${\cal I}^{(4)f'}$,
${\cal I}^{(4)e2'}$,
and
${\cal I}^{(4)c'}$
is left for future work.

\vskip 1cm

\centerline {\bf Acknowledgments}

We have benefited from discussions with
A.~Jevicki, Z.~Bern, D.~Kosower and J.~Maldacena and are grateful
to F.~Cachazo for collaboration in the early stages of this work.
The research of
MS is supported by NSF grant PHY-0610259 and by an OJI award
under DOE grant DE-FG02-91ER40688.
The research of AV is supported by NSF CAREER Award PHY-0643150 and
by DOE grant DE-FG02-91ER40688.

\smallskip

\listrefs
\end